\documentclass[twocolumn,showkeys,aps,prl,showpacs]{revtex4-1}
\usepackage{graphicx}
\usepackage[CJKbookmarks,dvipdfm,colorlinks,linkcolor=blue,citecolor=blue]{hyperref}

\begin{document}

\title{Pressure enhanced thermoelectric properties in $\mathrm{Mg_2Sn}$}

\author{San-Dong Guo and Jian-Li Wang}
\affiliation{Department of Physics, School of Sciences, China University of Mining and
Technology, Xuzhou 221116, Jiangsu, China}
\begin{abstract}
 Pressure dependence of electronic structures and  thermoelectric properties of  $\mathrm{Mg_2Sn}$ are investigated by using a modified Becke and Johnson (mBJ) exchange potential, including spin-orbit coupling (SOC). The corresponding  value  of spin-orbit splitting  at $\Gamma$ point is 0.47 eV,  which  is in good agreement with the experimental value 0.48 eV. With the pressure increasing,  the energy band gap first increases, and then decreases. In certain doping range, the power factor for n-type  has the same trend with energy band gap, when the  pressure increases.  Calculated results show that the pressure can lead to significantly enhanced power factor in  n-type doping below the critical pressure, and the corresponding lattice thermal conductivity near  the critical pressure shows the relatively small value.  These results  make us believe  that thermoelectric properties  of  $\mathrm{Mg_2Sn}$ can be improved  in n-type doping by pressure.

\end{abstract}


\maketitle

%\section{Introduction}

Thermoelectric material by using the Seebeck effect can convert waste heat directly to electricity  to solve energy problems.
The  performance  of thermoelectric material can be characterized by dimensionless  figure of merit\cite{s1,s2}, $ZT=S^2\sigma T/(\kappa_e+\kappa_L)$, where S, $\sigma$, T, $\kappa_e$ and $\kappa_L$ are the Seebeck coefficient, electrical conductivity, absolute  temperature, the electronic and lattice thermal conductivities, respectively.
Bismuth-tellurium systems\cite{s3,s4}, silicon-germanium alloys\cite{s5,s6}, lead chalcogenides\cite{s7,s8} and skutterudites\cite{s9,s10}  have been
identified as  excellent thermoelectric material for thermoelectric devices. For  thermoelectric research, the main objective is to   search for high $ZT$ materials, which has proven to be interesting and challenging.

The  thermoelectric material $\mathrm{Mg_2X}$(X = Si, Ge, Sn) composed of abundant, low-cost
elements and their alloys  have attracted much recent attention\cite{s11,s12,s13},
and various doping strategies have been adopted to attain high $ZT$\cite{s14,s15,s16}.
Pressure by tuning the electronic structures of materials   can accomplish many  interesting phenomenons like recent  pressure-induced high-Tc superconductivity in $\mathrm{(H_2S)_2H_2}$\cite{ht1,ht2}.
Here, we use first-principle calculations and Boltzmann transport theory to address the pressure dependence of thermoelectric
properties in the $\mathrm{Mg_2Sn}$.  Calculated results show that the  pressure dependence of energy band gap with mBJ+SOC
is consistent with one with mBJ\cite{gsd}, and first increases, and then decreases.  Pressure can  significantly improve  power factor in  n-type doping below the critical pressure. It is found that pressure can reduce the  lattice thermal conductivity in certain pressure range. These lead to enhanced $ZT$, and make $\mathrm{Mg_2Sn}$ become more efficient for thermoelectric application in n-type doping by pressure. So, pressure tuning offers a very effective method to  search
for materials with enhanced thermoelectric properties.

The rest of the paper is organized as follows. Firstly, we shall give our computational details.  Secondly, we shall present our main calculated results and analysis. Finally, we shall give our  discussion and conclusion.
\begin{figure}
  \includegraphics[width=8.0cm]{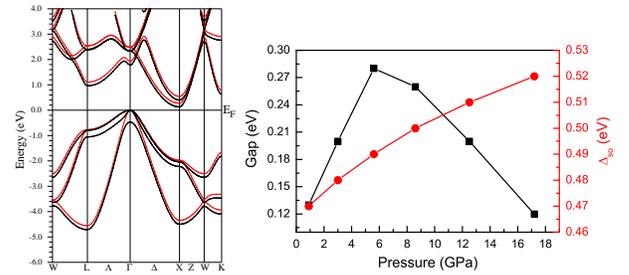}
  \caption{(Color online) Left: The energy band structures  by using mBJ (Red lines) and mBJ+SOC (Black lines). Right: The energy band gap (Gap) and  the value  of spin-orbit splitting at $\Gamma$ point ($\Delta_{so}$) as a function of pressure by using mBJ+SOC.}\label{t1}
\end{figure}

\begin{figure*}
  \includegraphics[width=15cm]{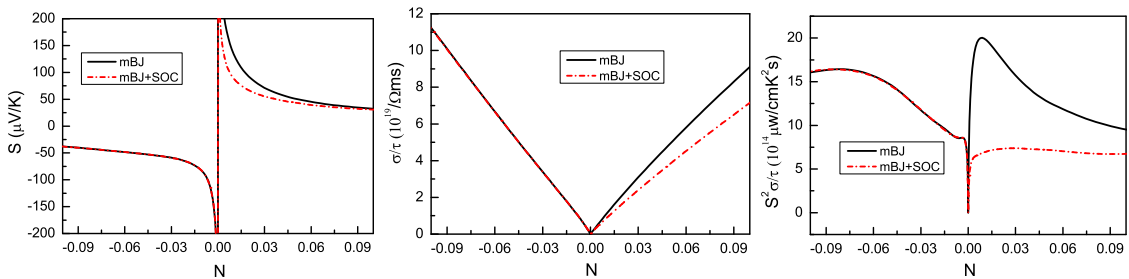}
  \caption{(Color online)  At temperature of 300 K,  transport coefficients  as a function of doping levels (electrons [minus value] or holes [positive value] per unit cell):  Seebeck coefficient S (Left), electrical conductivity with respect to scattering time  $\mathrm{\sigma/\tau}$ (Middle) and   power factor with respect to scattering time $\mathrm{S^2\sigma/\tau}$ (Right)  calculated with mBJ (Black solid line) and mBJ+SOC (Red dotted line).}\label{t2}
\end{figure*}
\begin{figure*}
  \includegraphics[width=15cm]{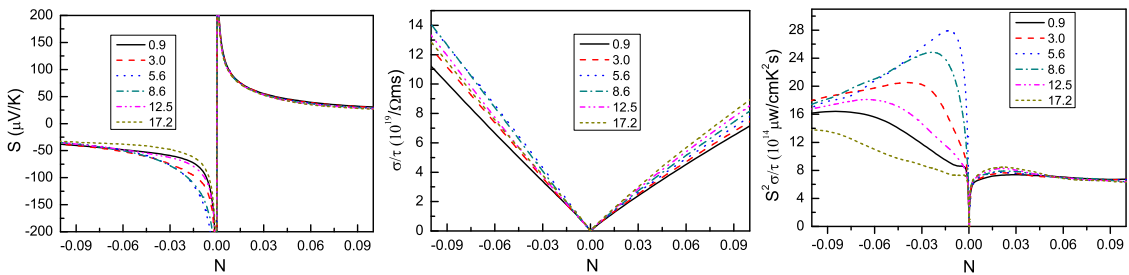}
  \caption{ At temperature of 300 K,  transport coefficients  as a function of doping levels (electrons [minus value] or holes [positive value] per unit cell):  Seebeck coefficient S (Left), electrical conductivity with respect to scattering time  $\mathrm{\sigma/\tau}$ (Middle) and   power factor with respect to scattering time $\mathrm{S^2\sigma/\tau}$ (Right)  with pressure being 0.9, 3.0, 5.6, 8.6, 12.5 and 17.2 (unit: GPa)  calculated by using mBJ+SOC. }\label{t3}
\end{figure*}
%\section{Computational detail}
We use a full-potential linearized augmented-plane-waves method
within the density functional theory (DFT) \cite{1}, as implemented in
the package WIEN2k \cite{2}.  We use Tran and Blaha's mBJ
 exchange potential plus local-density approximation (LDA)
correlation potential  for the
exchange-correlation potential \cite{4} to do our main DFT
calculations.  The full relativistic effects are calculated
with the Dirac equations for core states, and the scalar
relativistic approximation is used for valence states
\cite{10,11,12}. The SOC was included self-consistently
by solving the radial Dirac equation for the core electrons
and evaluated by the second-variation method\cite{so}. We use 6000 k-points in the
first Brillouin zone for the self-consistent calculation.
We make harmonic expansion up to $\mathrm{l_{max} =10}$ in each of the atomic spheres, and
set $\mathrm{R_{mt}*k_{max} = 8}$. The self-consistent calculations are
considered to be converged when the integration of the absolute
charge-density difference between the input and output electron
density is less than $0.0001|e|$ per formula unit, where $e$ is
the electron charge. Transport calculations
are performed through solving Boltzmann
transport equations within the constant
scattering time approximation as implemented in
BoltzTrap\cite{b}, which has been applied successfully to several
materials\cite{b1,b2,b3}. To
obtain accurate transport coefficients, we use 200000 k-points in the
first Brillouin zone for the energy band calculation. The  lattice thermal conductivities are calculated
by using Phono3py+VASP codes\cite{pv1,pv2,pv3,pv4}. For the third-order force constants, 2$\times$2$\times$2 supercells
are built, and reciprocal
spaces of the supercells are sampled by  2$\times$2$\times$2 meshes. To compute lattice thermal conductivities, the
reciprocal spaces of the primitive cells  are sampled using the 13$\times$13$\times$13 meshes.

%\section{MAIN CALCULATED RESULTS AND ANALYSIS}
The electronic structures, optical properties  and thermoelectric properties
of $\mathrm{Mg_2Sn}$ at hydrostatic pressure have been investigated by mBJ exchange-potential\cite{gsd}. However,
SOC is very important for power factor calculations\cite{w3}. Here, we investigate the electronic structures and thermoelectric properties
by using mBJ+SOC. First, the energy band structures of  $\mathrm{Mg_2Sn}$ with mBJ and mBJ+SOC are shown in \autoref{t1}. It is found that the SOC has little effect on the conduction bands, and has obvious influence on valence bands.  The SOC splits the valence band at $\Gamma$ point, and the corresponding  value  of spin-orbit splitting 0.47 eV is in good
agreement with the experimental value 0.48 eV\cite{spg}. As expected, the SOC reduces the energy band gap due to the conduction band minimum (CBM) moving toward lower energy. The energy band gap  and  value  of spin-orbit splitting at $\Gamma$ point  as a function of pressure by using mBJ+SOC  are present in \autoref{t1}. The trend of energy band gap with mBJ+SOC is consistent with one with mBJ\cite{gsd}, and first increases, and then decreases with increasing pressure. The explanation of
  trend of energy band gap can be found in ref.\cite{gsd}. But, the spin-orbit splitting monotonically increases.
\begin{figure}
  \includegraphics[width=7cm]{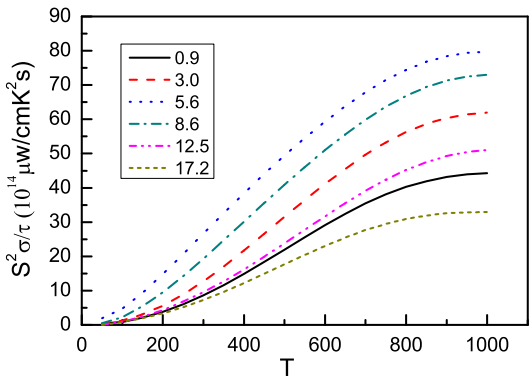}
  \caption{(Color online) Power factor with respect to scattering time $\mathrm{S^2\sigma/\tau}$ as a function of temperature for n-type  with pressure being 0.9, 3.0, 5.6, 8.6, 12.5 and 17.2 (unit: GPa)  calculated by using mBJ+SOC  with the doping concentration of $\mathrm{1\times10^{20}cm^{-3}}$.}\label{t5}
\end{figure}

 $\mathrm{Mg_2Sn}$ based thermoelectric materials are considered as  potential
candidates for efficient thermoelectricity. The  pressure dependence of   the  semi-classic transport coefficients as a function
of doping level is investigated  within constant scattering time approximation Boltzmann theory.
 We firstly consider the SOC effects on  the Seebeck coefficient S,   electrical conductivity with respect to scattering time  $\mathrm{\sigma/\tau}$ and  power factor with respect to scattering time $\mathrm{S^2\sigma/\tau}$,   and they  as  a function of doping levels  at the temperature of 300 K by using mBJ and mBJ+SOC are present in \autoref{t2}.
  It is clearly seen that the negative doping levels (n-type doping)  show  the negative Seebeck coefficient, and the positive doping levels (p-type doping) imply the positive Seebeck coefficient.
Calculated results show that SOC has a detrimental influence on S , $\mathrm{\sigma/\tau}$ and $\mathrm{S^2\sigma/\tau}$ in p-type doping, but has a negligible effect in n-type doping.
These can be explained  by  that SOC has larger influences on the valence bands  than on the conduction bands.
When SOC is absent,  the best  p-type   power factor  is larger than the best n-type one. However, including SOC, the power factor in n-type doping  is larger than one in p type doping in considered doping range, which agrees with the experimental results reporting  high  $ZT$ values  for n-type than for p-type\cite{s14}.

\begin{figure}
  \includegraphics[width=7cm]{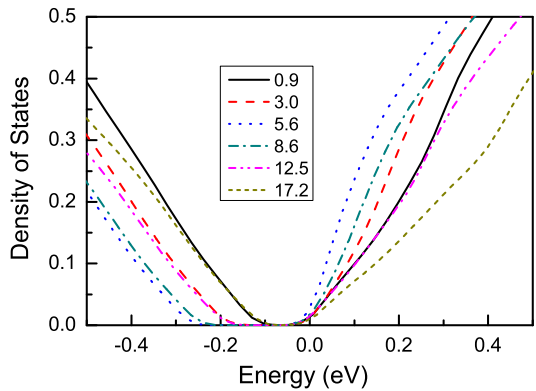}
  \caption{(Color online) The  total density of states   with pressure being 0.9, 3.0, 5.6, 8.6, 12.5 and 17.2 (unit: GPa)  calculated by using mBJ+SOC. The Fermi level is defined at the conduction band minimum (CBM).}\label{t6}
\end{figure}

The pressure dependence of S,  $\mathrm{\sigma/\tau}$ and $\mathrm{S^2\sigma/\tau}$    with pressure being 0.9, 3.0, 5.6, 8.6, 12.5 and 17.2 (unit: GPa) calculated by using mBJ+SOC at temperature of 300 K are shown in \autoref{t3}. It is interesting that
S (absolute value), $\mathrm{\sigma/\tau}$ and $\mathrm{S^2\sigma/\tau}$  have the same trend with energy band gap in certain doping range for n-type with the increasing pressure.  When the pressure reaches the critical value  of energy band gap, the S,  $\mathrm{\sigma/\tau}$ and   $\mathrm{S^2\sigma/\tau}$ attain  the corresponding extremum.
For p-type, the  $\mathrm{\sigma/\tau}$ has obvious dependence on pressure, but the S has very weak dependence on pressure, which leads to weak  pressure dependence for   $\mathrm{S^2\sigma/\tau}$. The strong  pressure dependence for   $\mathrm{S^2\sigma/\tau}$ in n-type doping shows that the $\mathrm{Mg_2Sn}$ under pressure may  become more efficient thermoelectric material. To clearly see interesting  pressure dependence in n-type doping, the $\mathrm{S^2\sigma/\tau}$ as a function of temperature  with the doping concentration of  $\mathrm{1\times10^{20}cm^{-3}}$ are displayed in \autoref{t5}.
In the considered temperature range, the power factor  always has the same trend with energy band gap.

To explain interesting pressure dependence of  power factor in n-type doping, the total density of states with pressure being 0.9, 3.0, 5.6, 8.6, 12.5 and 17.2 (unit: GPa)  calculated by using mBJ+SOC are displayed in \autoref{t6}.  Calculated results imply that the slope of density of states of the conduction bands near the energy band gap first increases  with the pressure  increasing, and then decreases. The critical pressure happens to be the critical one for power factor.
The large slope of density of states near the energy gap may
induce a large Seebeck coefficient in narrow-gap semiconductors\cite{hf1}, leading to large power factor, which gives rise to the corresponding pressure dependence of power factor in n-type doping.

\begin{figure}
  % Requires \usepackage{graphicx}}
  \includegraphics[width=7cm]{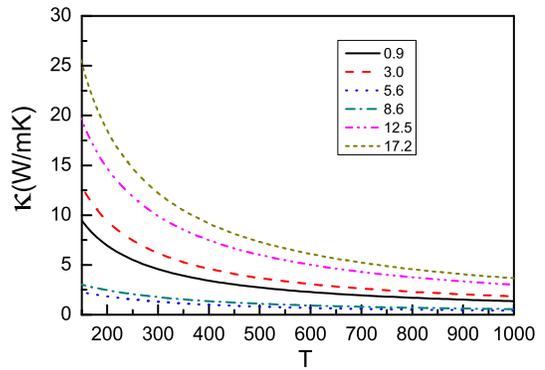}
  \caption{(Color online) The lattice thermal conductivities $\kappa$ as a function of temperature   with pressure being 0.9, 3.0, 5.6, 8.6, 12.5 and 17.2 (unit: GPa)  calculated by using GGA.}\label{t7}
\end{figure}

Finally, the lattice thermal conductivities $\kappa$ as a function of temperature   with pressure being 0.9, 3.0, 5.6, 8.6, 12.5 and 17.2 calculated by using GGA are shown in \autoref{t7}. The lattice thermal
conductivity is generally   considered to be  independent of  doping, and
 typically goes as 1/T.  At high T , it can reach the so-called minimum thermal
conductivity. Calculated results show that the $\kappa$ near  the critical pressure of energy band gap  has the relatively small value, and the corresponding  power factor in n-type doping has the relatively large one. These results imply that pressure can induce larger $ZT$ by reducing lattice thermal conductivity and enhancing power factor for n-type.

%\section{Discussions and Conclusion}
It has been proved that mBJ gap value agrees well with experimental value 0.3 eV\cite{gsd}, but mBJ+SOC gap value is less than experimental one. However, the mBJ+SOC is  more satisfactory than usual GGA or  LDA+SOC in calculating electronic structure of  $\mathrm{Mg_2Sn}$. It is very important for power factor calculations  to consider SOC for MgSn, especially for p-type doping.  When SOC is included, the n-type doping has more excellent power factor than p-type doping, which agrees that  the best p-type material reported so far has  lower $ZT$ than the best n-type.  Pressure has obvious effects on the conduction bands, and has little influences on valence bands, which leads to  remarkable effects on power factor for n-type and small effects in p-type doping.

In summary,  we investigate pressure dependence of  thermoelectric properties  of $\mathrm{Mg_2Sn}$ by using mBJ+SOC .  It is found that pressure  can realize enhanced power factor below critical pressure and reduced   lattice thermal conductivity near the critical pressure, which leads to the improved $ZT$ for efficient thermoelectric application.
By choosing the  appropriate doping concentration,  $\mathrm{Mg_2Sn}$ under pressure can provide great opportunities for efficient thermoelectricity.

\begin{acknowledgments}
This work is supported supported by the Fundamental Research Funds for the Central Universities (2015XKMS073). We are grateful to the Advanced Analysis and Computation Center of CUMT for the award of CPU hours to accomplish this work.
\end{acknowledgments}


\begin{references}
\bibitem{s1} Y. Pei, X. Shi, A. LaLonde, H. Wang, L. Chen and G. J. Snyder, Nature \textbf{473}, 66 (2011).

\bibitem{s2} A. D. LaLonde, Y. Pei, H. Wang and G. J. Snyder, Mater. Today \textbf{14}, 526 (2011).
\bibitem{s3} W. S. Liu, Q. Y. Zhang, Y. C. Lan, S. Chen, X. Yan, Q. Zhang, H. Wang,
D. Z. Wang, G. Chen and Z. F. Ren, Adv. Energy Mater. \textbf{1},  577 (2011).

\bibitem{s4}D. K. Ko, Y. J. Kang and C. B. Murray, Nano Lett.,  \textbf{11}, 2841 (2011).

\bibitem{s5}M. Zebarjadi, et al. Nano Lett.  \textbf{11},  2225 (2011).

\bibitem{s6}B. Yu, et al. Nano Lett.  \textbf{12},  2077 (2012).


\bibitem{s7}Y. Z. Pei, X. Y. Shi, A. Lalonde  et al, Nature \textbf{473}, 66  (2011).

\bibitem{s8}J. Q. He, J. R. Sootsman, S. N. Girard  et al,  J. Am. Chem. Soc. \textbf{132}, 8669  (2010).


\bibitem{s9} A. C. Sklad, M. W. Gaultois and A. P. Grosvenor,   J. Alloys Compd. \textbf{505}, L6 (2010).

\bibitem{s10}X. Shi, J. Yang and J. R. Salvador, J. Am. Chem. Soc. \textbf{133}, 7837 (2011).


\bibitem{s11}W. Liu, X. Tan, K. Yin, H. Liu, X. Tang, J. Shi, Q. Zhang, and
C. Uher, Phys. Rev. Lett. \textbf{108}, 166601 (2012).

\bibitem{s12}W. J. Luo, M. J. Yang, Q. Shen, H. Y. Jiang and L. Zhang, Adv.
Mater. Res. \textbf{66}, 33 (2009).

\bibitem{s13}M. Yang, W. Luo, Q. Shen, H. Jiang and L. Zhang, Adv. Mater.
Res. \textbf{66}, 17 (2009).

\bibitem{s14}V. K. Zaitsev, M. I. Fedorov, E. A. Gurieva, I. S. Eremin, P. P.
Konstantinov, A. Y. Samunin and M. V. Vedernikov, Phys. Rev. B
 \textbf{74}, 045207 (2006).

\bibitem{s15}Q. Zhang, J. He, T. J. Zhu, S. N. Zhang, X. B. Zhao and T. M. Tritt,
Appl. Phys. Lett.  \textbf{93}, 102109 (2008).

\bibitem{s16}W. Liu, X. Tang and J. Sharp, J. Phys. D: Appl. Phys.  \textbf{43}, 085406
(2010).

\bibitem{ht1}D. F. Duan, Y. X.  Liu, F. B. Tian, D. Li, X. L. Huang, Z. L. Zhao, H. Y.  Yu, B. B. Liu, W. J. Tian  and  T. Cui,
Sci. Rep.  \textbf{4}, 6968  (2014).

\bibitem{ht2}A. P. Drozdov,	 M. I. Eremets,	 I. A. Troyan,	 V. Ksenofontov	 and  S. I. Shylin, Nature \textbf{525}, 73 (2015).

\bibitem{gsd}S. D. Guo, EPL \textbf{109},  57002 (2015).

%%%%%%%%%%%%%%%%%%%%%%%%%%%%%%%%%%%%%%%%%%%%%%%%%%%%%%%%%%%%%%%%%%%%
\bibitem{1}P. Hohenberg and W. Kohn, Phys. Rev. \textbf{136},
B864 (1964); W. Kohn and L. J. Sham, Phys. Rev. \textbf{140},
A1133 (1965).

\bibitem{2}P. Blaha, K. Schwarz, G. K. H. Madsen, D. Kvasnicka
 and J. Luitz, WIEN2k, an Augmented Plane Wave
+ Local Orbitals Program for Calculating Crystal Properties
(Karlheinz Schwarz Technische Universit\"at Wien, Austria) 2001,
ISBN 3-9501031-1-2
\bibitem{4}F. Tran and P. Blaha, Phys. Rev. Lett. \textbf{102},
226401 (2009).

%\bibitem{pbe}J. P. Perdew, K. Burke and M. Ernzerhof, Phys. Rev. Lett. \textbf{77}, 3865 (1996).

\bibitem{10}A. H. MacDonald, W. E. Pickett and D. D. Koelling, J. Phys. C \textbf{13}, 2675 (1980).

\bibitem{11}D. J. Singh and L. Nordstrom, Plane Waves, Pseudopotentials and the LAPW
Method, 2nd Edition (Springer, New York, 2006).

\bibitem{12}J. Kunes, P. Novak, R. Schmid, P. Blaha and
K. Schwarz, Phys. Rev. B \textbf{64}, 153102 (2001).

\bibitem{so}D. D. Koelling, B. N. Harmon, J. Phys. C Solid State Phys.  \textbf{10}, 3107 (1977).



\bibitem{b}G. K. H. Madsen and D. J. Singh, Comput. Phys. Commun. \textbf{175}, 67
(2006).

\bibitem{b1}B. L. Huang and M. Kaviany, Phys. Rev. B \textbf{77}, 125209 (2008).

\bibitem{b2}L. Q. Xu, Y. P. Zheng and J. C. Zheng, Phys. Rev. B \textbf{82}, 195102 (2010).

\bibitem{b3}J. J. Pulikkotil, D. J. Singh, S. Auluck, M. Saravanan, D. K. Misra, A. Dhar and R. C. Budhani,
Phys. Rev. B \textbf{86}, 155204 (2012).

\bibitem{pv1} G. Kresse, J. Non-Cryst. Solids \textbf{193}, 222 (1995).

\bibitem{pv2} G. Kresse and J. Furthm¡§uller, Comput. Mater. Sci. 6, \textbf{15} (1996).

\bibitem{pv3} G. Kresse and D. Joubert, Phys. Rev. B \textbf{59}, 1758 (1999).

\bibitem{pv4}A. Togo, L. Chaput and I. Tanaka, Phys. Rev. B \textbf{91}, 094306 (2015).

%%%%%%%%%%%%%%%%%%%%%%%%%%%%%%%%%%%%%%%%%%%%%%%%%%%%%%%%%%%%%%%%%%%%%%%%%%%%%%%%%%%%%%%%%%%%%
\bibitem{w3}K. Kutorasinski, B. Wiendlocha, J. Tobola and S. Kaprzyk,
Phys. Rev. B \textbf{89}, 115205 (2014).


\bibitem{spg}F. Vazquez, A. R. Forman and M. Cardonna, Phys. Rev. \textbf{176},
905 (1968).

\bibitem{hf1} M. ONOUE, F. ISHII and T. OGUCHI, J. Phys. Soc. Jpn.  \textbf{77}, 054706 (2008).

\end{references}
\end{document}